# Basic mechanisms of escape of a harmonically forced classical particle from a potential well.

*O.V.Gendelman\* and G. Karmi*

Technion – Israel Institute of Technology,

Haifa, 3200003, ISRAEL

\* - corresponding author, ovgend@technion.ac.il

Tel. +97248293877, ORCID: 0000-0002-4750-2708

In various models and systems involving the escape of periodically forced particle from the potential well, a common pattern is observed. Namely, the minimal forcing amplitude required for the escape exhibits sharp minimum for the excitation frequency below the natural frequency of small oscillations in the well. The paper explains this regularity by exploring the transient escape dynamics in simple benchmark potential wells. In the truncated parabolic well, in absence of the damping the minimal forcing amplitude obviously tends to zero for the natural excitation frequency. Addition of weak symmetric softening nonlinearity to the truncated parabolic well leads to the nonzero forcing minimum below the natural frequency. We explicitly compute this shift in the principal approximation by considering the slow-flow dynamics in conditions of the principal 1:1 resonance. Essentially nonlinear $\varphi^4$ model, analyzed with the help of transformation to action-angle variables, demonstrates very similar qualitative features of the transient escape dynamics.

## 1. Introduction

The concepts of potential well, and escape from it, are crucial in physics, chemistry and engineering [1-5]. Very incomplete list of research topics related to the escape processes includes dynamics of molecules and absorbed particles, celestial mechanics and gravitational collapse, energy harvesting [6], physics of Josephson junctions [7], transient resonance dynamics of oscillatory systems [8, 9], and even such deceivingly remote topic as capsize of ships [3, 10]. Another widely explored and useful engineering phenomenon governed by the escape dynamics is a dynamic pull-in in microelectromechanical systems (MEMS) [11-14].

Historically, the exploration of the forced escape started in $1940^{th}$, from famous works of Kramers devoted to thermal activation of chemical reactions [15, 16]. Despite more than 70 years of active development, this research area remains active and vibrant, and contains a substantial list of unresolved issues [17]. Among many startling phenomena related to the stochastically activated escape, one encounters widely explored stochastic resonances [18, 19].

In very different situation, when the forcing is constant, the escape can occur because of slow variation of the system parameters and subsequent bifurcations of the steady-state response regimes [2]. Paper [4] considers escape of the particle under harmonic forcing with constant amplitude and frequency, and with zero initial conditions (IC), from three different potential wells. Analytic treatment adopted by the authors involves evaluation of possible steady-state responses of the particle. The escape is associated with energy of the response that exceeds the potential barrier. Empirical correction factors are introduced, if required.

All model systems explored in [4] exhibit a peculiar common feature. The critical force amplitude required for the escape demonstrates a sharp minimum at certain frequency below the frequency of small oscillations in the well. Qualitatively similar escape curves in frequency – voltage domain were observed in the problem of dynamic pull-in in MEMS [12,13]. Exploration of safe basins of attraction for various dynamical models with possibility of escape revealed somewhat similar critical patterns [20, 21]. With appropriate caution, it is possible to conjecture that the aforementioned phenomenological feature – sharp minimum of the critical force at certain frequency below the natural – is a "fingerprint" of the escape process in conditions of the periodic forcing.

This conjecture was further corroborated in recent detailed exploration of the escape from special (square of hyperbolic secant) potential well under the harmonic forcing [22]. It was demonstrated that one could successfully explain the escape dynamics in this model system by considering the 1:1 resonance manifold (RM) - the phase portrait for the slow flow of the system. The escape on the RM is predicted by tracking the behavior of the special phase trajectory, determined by the IC. For zero IC, such special trajectories are sometimes referred to as limiting phase trajectories (LPT, [23, 24]). It was demonstrated that the escape thresholds are related to passage of these special trajectories through saddle points of the RM. The slow-flow equations exhibit two saddle fixed points, and therefore, two competing escape mechanisms on the RM. One can assume that this competition, in turn, leads to this peculiar sharp minimum at the critical forcing vs. frequency plot. Similar mechanism holds also in the presence of viscous damping [25].

Despite the aforementioned efforts, it is still not clear what physical reasons lead to the observed common phenomenology of the escape process under harmonic forcing. The paper attempts to clarify this issue. First, we address arguably the simplest possible model of particle in truncated parabolic well with harmonic forcing (Section 2). Quite obviously, the sharp minimum at zero critical forcing appears in this model for the natural forcing frequency (conditions of exact resonance). In Section 3 we perturb the parabolic truncated well by adding small softening quartic nonlinearity. Due to simplicity of the perturbed model, it is possible to calculate explicitly the critical forcing. It turns out that the increasing perturbation shifts the minimum upwards (to higher minimal forcing amplitude) and leftwards (to lower frequency). Similar qualitative behavior and similar basic structure of the RM near 1:1 resonance are revealed in Section 4 also for fully nonlinear quadratic ($\varphi^4$) model. This Section is followed by conclusions.

## 2. The simplest benchmark model: truncated parabolic potential well

In the model of truncated parabolic potential well, we consider a particle of mass $m$ with the following potential energy:

$$V(x) = \begin{cases} -V_0 + \dfrac{m\omega_0^2}{2} x^2, & |x| \leq \dfrac{1}{\omega_0}\sqrt{\dfrac{2V_0}{m}} \\ 0, & |x| > \dfrac{1}{\omega_0}\sqrt{\dfrac{2V_0}{m}} \end{cases} \qquad (1)$$

Here $V_0$ is the depth of the potential well, $\omega_0$ is the frequency of free oscillations in the well, and $x(t)$ is physical displacement of the particle. Equation of motion of the particle is written as follows:

$$m\frac{d^2 x}{dt^2} + \frac{\partial V}{\partial x} = A\sin(\omega t + \Psi) \qquad (2)$$

A is the amplitude of the external forcing, $\omega$ - its frequency, and $\Psi$ - the forcing phase. The escape problem naturally considers the transient responses, and IC of the particle have critical significance. Therefore, it is not possible to "remove" the forcing phase $\Psi$ by re-definition of the time, and it constitutes a significant parameter of the problem.

In terms of non-dimensional variables, the problem is reformulated in the following way:

$$\ddot{q} + \frac{\partial U_0}{\partial q} = F\sin(\Omega\tau + \Psi), \quad U_0(q) = \begin{cases} -\dfrac{1}{2} + \dfrac{q^2}{2}, & |q| \leq 1 \\ 0, & |q| > 1 \end{cases}$$

$$\tau = \omega_0 t, \; q = x\omega_0\sqrt{\dfrac{m}{2V_0}}, \; \Omega = \dfrac{\omega}{\omega_0}, \; F = \dfrac{A}{\omega_0\sqrt{2mV_0}} \qquad (3)$$

The dot here and in the following text denotes differentiation with respect to the non-dimensional time τ. Sketch of the parabolic truncated potential well $U_0(q)$ is presented in Figure 1. From Equation (3) it is clear that the escape of the particle from this well is governed by the IC and the values of non-dimensional forcing $F$, non-dimensional frequency $\Omega$ and the forcing phase $\Psi$.

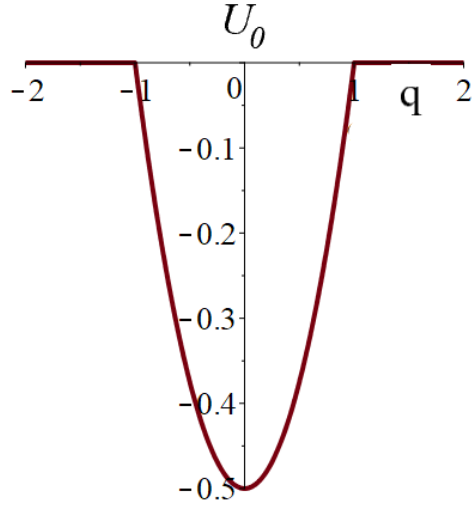

Figure 1. Sketch of the parabolic truncated potential well (cf. Equation (3))

The particle escapes the well if $\lim_{\tau \to \infty}|q(\tau)| > 1$. Practically, this criterion is not very convenient for implementation, and we use the alternative one for the escape threshold:

$$\max_\tau |q(\tau)| = 1 \tag{4}$$

Formally, these two criteria are not equivalent, but for the considered zero IC $q(0) = 0, \dot{q}(0) = 0$ the difference turns out negligible. For given IC, Equation (3) is easily solved assuming $|q(\tau)| < 1$, and the solution takes the following form:

$$q(\tau) = \frac{F}{1-\Omega^2}\left(\sin(\Omega\tau + \Psi) - \Omega \sin\tau \cos\Psi - \cos\tau \sin\Psi\right) =$$
$$= \frac{F}{1-\Omega^2}\left(\sin(\Omega\tau + \Psi) - \sqrt{\Omega^2 \cos^2\Psi + \sin^2\Psi}\,\sin(\tau + \varphi)\right); \tag{5}$$
$$\tan\varphi = \frac{1}{\Omega}\tan\Psi$$

In accordance with (4-5), the threshold forcing for the escape for given excitation frequency $\Omega$ and forcing phase $\Psi$, denoted as $F_{tr}(\Omega, \Psi)$, is determined by the following equation:

$$F_{tr}(\Omega, \Psi) = \frac{|1-\Omega^2|}{\max_{\tau}\left|\sin(\Omega\tau+\Psi) - \sqrt{\Omega^2\cos^2\Psi + \sin^2\Psi}\sin(\tau+\varphi)\right|} \tag{6}$$

The truncated parabolic potential is simple enough to allow explicit derivation of the exact closed-form expression (6) for the escape threshold. However, despite all simplifications, evaluation of the maximum in the denominator leads to a nontrivial transcendent problem with, generally speaking, no possibility of exact solution for arbitrary combinations of $(\Omega, \Psi)$. To treat this problem, we first note the obvious inequality

$$\max_{\tau}\left|\sin(\Omega\tau+\Psi) - \sqrt{\Omega^2\cos^2\Psi + \sin^2\Psi}\sin(\tau+\varphi)\right| \leq 1 + \sqrt{\Omega^2\cos^2\Psi + \sin^2\Psi} \tag{7}$$

Then, one obtains the following estimation:

$$F_{tr}(\Omega, \Psi) \geq F_c(\Omega, \Psi) = \frac{|1-\Omega^2|}{1 + \sqrt{\Omega^2\cos^2\Psi + \sin^2\Psi}} \tag{8}$$

Inequality (8) yields the lower boundary for $F_{tr}(\Omega, \Psi)$ in all parametric range $\Omega > 0, 0 \leq \Psi \leq \pi$. Other values of $\Psi$ do not require treatment due to the symmetry of the problem. Inequalities (7) and (8) become equalities when sine functions in Equation (6) achieve values 1 and -1 simultaneously. Mathematically, the following conditions should be fulfilled for some time instance $t_m$ and certain integers $(k,l,m)$:

$$\Omega t_m + \Psi = \frac{\pi}{2}(2k+1), \quad t_m + \varphi = \frac{\pi}{2}(2l+1) \tag{9}$$
$$k, l \in \mathbb{Z}, \quad k+l = 2m+1, \quad m \in \mathbb{Z}$$

The last condition in (9) follows from minus sign in (7) – numbers $k$ and $l$ should have different parity. Excluding $t_m$ from (9), one obtains:

$$\Omega(2l+1) - \frac{2\Omega\varphi}{\pi} + \frac{2\Psi}{\pi} - 2k - 1 = 0 \tag{10}$$

Therefore, Equations (9) with all conditions will satisfied if the straight line on $(x,y)$ plane, described by the equation

$$\Omega(2x+1) - \frac{2\Omega\varphi}{\pi} + \frac{2\Psi}{\pi} - 2y - 1 = 0 \qquad (11)$$

will pass through points with integer coordinates of different parity:

$$(x, y) = (k, l), \ k, l \in \mathbb{Z}, \ k + l = 2m + 1, \ m \in \mathbb{Z} \qquad (12)$$

In order to further explore this problem, one can map the complete $(x,y)$ plane to torus:

$$x \mapsto x \bmod 2, \ y \mapsto y \bmod 2 \qquad (13)$$

Mapping (13) maps all points with integer coordinates of opposite parity into points $(0,1)$ or $(1,0)$, and the map of straight line (11) winds the torus. As it is well-known, if $\Omega$ is irrational number, this winding is dense everywhere, and the winding trajectory either passes through the required points, or passes arbitrarily close to both of them. By continuity, in all these cases it is possible to conclude that for all *irrational* values of the forcing

$$F_{tr}(\Omega, \Psi) = F_c(\Omega, \Psi) = \frac{|1-\Omega^2|}{1+\sqrt{\Omega^2 \cos^2 \Psi + \sin^2 \Psi}}, \Omega \in \{0, \mathbb{R}^+\}/\mathbb{Q}, \ 0 \le \Psi \le \pi \qquad (14)$$

For rational values of $\Omega$, the winding trajectory on the torus (13) will be periodic, and therefore for fixed $\Omega$ it will either pass through the points $(0,1)$ and/or $(1,0)$ for some special values of $\Psi$, or will remain at certain nonzero minimal distance from these points. In this latter case the inequality in (7) will become strict, and, consequently, $F_{tr} > F_c$. One can already see that the dependence $F_{tr}(\Omega, \Psi)$ is discontinuous, even in this oversimplified model of the potential well. This discontinuity apparently stems from the discontinuity of the force in Equation (3).

As particular examples, one can easily derive the following identities for interesting particular cases of $\Psi = 0$ and $\Psi = \pi/2$:

$$F_{tr}(3,0) = F_c(3,0) = 3/2;\ F_{tr}(1/3,0) = F_c(1/3,0) = 4/9;$$

$$F_{tr}(2,0) = \frac{96}{\sqrt{30+2\sqrt{33}}\left(\sqrt{33}+3\right)} \approx 1.70 > F_c(2,0) \tag{15}$$

$$F_{tr}(1/2,0) = \frac{24}{\sqrt{30+2\sqrt{33}}\left(\sqrt{33}+3\right)} \approx 0.426 > F_c(1/2,0)$$

$$F_{tr}(3,\pi/2) = 2\sqrt{2} > F_c(3,\pi/2);\ F_{tr}(1/3,\pi/2) = \frac{2\sqrt{2}}{3} > F_c(1/3,\pi/2);$$
$$F_{tr}(2,\pi/2) = F_c(2,\pi/2) = 1;\ F_{tr}(1/2,\pi/2) = F_c(1/2,\pi/2) = 1/2; \tag{16}$$

One can see that for some rational values of the forcing frequency the real forcing threshold $F_{tr}(\Omega,\Psi)$ can substantially exceed the estimation $F_c(\Omega,\Psi)$ that is *exact for almost all* values of $\Omega$. It is tempting to identify these special values of the forcing frequency with *subharmonic or superharmonic resonances* [1, 26], but strong caution should be taken here. Commonly, when considering these resonances, one looks for steady-state responses or phase trajectories in the vicinity of such responses [26]. In current problem, however, the situation is opposite – the escape response is intrinsically transient, moreover, the escape happens only once. Therefore, the question of relationship between these special frequency values and subharmonic/superharmonic resonances requires special exploration, and is beyond the scope of this paper.

Instead, we will focus our attention at the case of main *1:1 resonance*, since it is the most interesting, and the most important in the applications. This resonance is characterized by closeness between the forcing frequency and the natural frequency of the oscillator.

It is clear both from (6) and (8) that both $F_{tr}$ and $F_c$ tend to zero as $\Omega \to 1$. In this limit, setting $\Omega = 1+\delta, |\delta| \ll 1$, one obtains from (8):

$$F_c(\delta,\Psi) = \begin{cases} \delta + \dfrac{\delta^2}{2}\sin^2\Psi + O(\delta^3), \delta > 0 \\ -\delta - \dfrac{\delta^2}{2}\sin^2\Psi + O(\delta^3), \delta < 0 \end{cases} \tag{17}$$

Expression (17) demonstrates that the critical forcing amplitude in the case of the main resonance does not depend on the forcing phase in the main approximation. To illustrate this result (Figure 2), we plot the expressions for $F_c$ as function of $\Omega$ for two particular values $\Psi = 0, \pi/2$:

$$F_c = |1-\Omega|, \Psi = 0$$
$$F_c = \frac{|1-\Omega^2|}{2}, \Psi = \frac{\pi}{2} \tag{18}$$

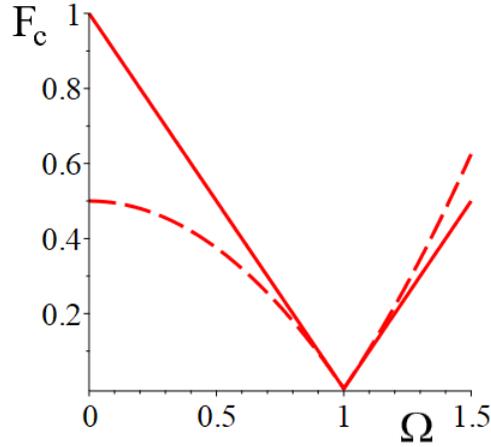

*Figure 2. Critical forcing amplitude $F_c(\Omega, \Psi)$ for $\Psi = 0$ (solid line) and $\Psi = \pi/2$ (dashed line).*

In this Figure, one observes the sharp minimum (in this case, zero) of the critical forcing at $\Omega \to 1$. One can conjecture that this sharp "main resonance dip" in the model of truncated parabolic well is in fact the prototype solution for the family of typical observed escape patterns, as mentioned in the Introduction. In other terms, all observed sharp minima of the critical forcing at least at qualitative level can be considered as "perturbations" of this basic solution. To elaborate this idea, we consider the effect of weak nonlinear perturbation of the truncated parabolic well on the main resonance dip.

3. **Escape from weakly nonlinear truncated well in conditions of main resonance.**

*3.1 Description of the model.*

In this Section, we consider the escape problem for the truncated parabolic well, weakly perturbed by softening quartic nonlinearity. Equation (3) is modified to the following form:

$$\ddot{q} + \frac{\partial U_\varepsilon}{\partial q} = F\sin(\Omega\tau + \Psi), \ U_\varepsilon(q) = \begin{cases} -\frac{1}{2} + \frac{q^2}{2} - \frac{\varepsilon\alpha q^4}{4}, |q| \leq q_m \\ 0, \ |q| > q_m \end{cases},$$

$$0 < \varepsilon \ll 1, \alpha > 0, q_m = \sqrt{\frac{1 - \sqrt{1 - 2\varepsilon\alpha}}{\varepsilon\alpha}}$$

(19)

Parameter $\varepsilon$ determines the strength of the perturbation for all involved terms, parameter $\alpha > 0$ governs the strength of the softening nonlinearity. The half-width of the well $q_m$ is chosen to ensure continuity of the potential function. The truncated potential well $U_\varepsilon(q)$ is illustrated in Figure 3.

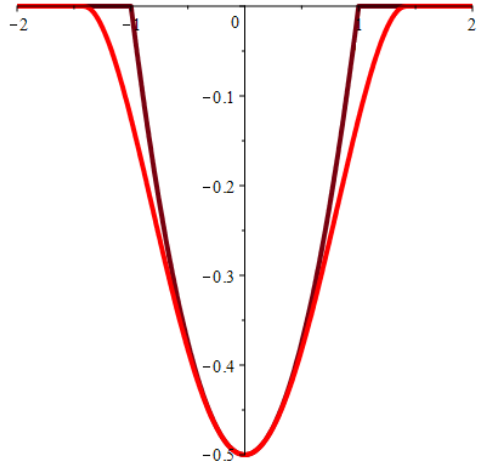

*Figure 3. Truncated parabolic well with softening quartic perturbation, $\varepsilon = 0.1, \alpha = 5$. Brown – unperturbed well $U_0(q)$, red - perturbed well $U_\varepsilon(q)$.*

*3.2 Averaging-based analysis.*

Contrary to Equation (3), it is not possible to elaborate the Equation (19) exactly even inside the well. As mentioned above, here we are interested primarily in the regime of main 1:1 resonance. One can conveniently analyze the transient dynamics near the 1:1 resonance *inside the potential well* in the lowest-order approximation with the help of primary averaging [26, 27, 28]. This procedure can be shaped in many forms, but here for the sake of generality we use the action-angle (AA) formalism [1, 27, 28, 29]. It is briefly presented here for the sake of completeness, based on paper [22].

System (19) can be derived from the following Hamiltonian:

$$H = H_0(p,q) - Fq\sin(\Omega\tau + \Psi); \quad H_0 = \frac{p^2}{2} + U_\varepsilon(q), \quad p = \dot{q}. \tag{20}$$

$H_0(p,q)$ is the component of the Hamiltonian that describes free motion of the particle in the potential well. Then, we perform a transformation to the AA variables in accordance with well-known formulas [1, 28] for certain basic Hamiltonian $H_b(p,q)$:

$$H_b(p,q) = E = \text{const}$$
$$I(E) = \frac{1}{2\pi}\oint p(q,E)dq; \quad \theta = \frac{\partial}{\partial I}\int_0^q p(q,I)dq \tag{21}$$

Here $H_b(p,q) = E = \text{const}$ defines a constant energy level for *some* basic Hamiltonian $H_b(p,q)$, *not necessarily equal* to $H_0(p,q)$. By inverting expressions (21), one can obtain explicit formulas for the canonical change of variables $p = p(I,\theta), q = q(I,\theta)$. The canonical transformation defined hereby does not include the explicit time dependence; therefore, the Hamiltonian of System (20) is written in the following form in terms of the AA variables:

$$H = H_0(I,\theta) - Fq(I,\theta)\sin(\Omega\tau + \Psi). \tag{22}$$

If the conservative part of Hamiltonian (20) would be used for the AA transformation (i.e. if $H_0(p,q) = H_b(p,q)$), then the transformed $H_0$ would not depend on angle variable θ; however, we prefer to maintain more general framework. In any case, due to 2π-periodicity of the angle variable, Hamiltonian (22) can be rewritten in terms of Fourier series [22, 27, 29]:

$$H = \sum_{m=-\infty}^{\infty} h_m(I)\exp(im\theta) + \frac{iF}{2}\sum_{m=-\infty}^{\infty} q_m(I)\left[\exp i(m\theta + \Omega\tau + \Psi) - \exp i(m\theta - \Omega\tau - \Psi)\right] \tag{23}$$
$$h_m = h_{-m}^*, \quad q_m = q_{-m}^*$$

Then, Hamilton equations will take the form:

$$\dot{I} = -\frac{\partial H}{\partial \theta} = -i \sum_{m=-\infty}^{\infty} m h_m(I) \exp(im\theta) + \frac{F}{2} \sum_{m=-\infty}^{\infty} m q_m(I) \left[ \exp i(m\theta + \Omega\tau + \Psi) - \exp i(m\theta - \Omega\tau - \Psi) \right]$$

$$\dot{\theta} = \frac{\partial H}{\partial I} = \sum_{m=-\infty}^{\infty} \frac{\partial h_m(I)}{\partial I} \exp(im\theta) + \frac{iF}{2} \sum_{m=-\infty}^{\infty} \frac{\partial q_m(I)}{\partial I} \left[ \exp i(m\theta + \Omega\tau + \Psi) - \exp i(m\theta - \Omega\tau - \Psi) \right]$$

(24)

To treat the regime of 1:1 resonance, one should *assume* slow evolution of the phase variable $\vartheta = \theta - \Omega\tau - \Psi$; all other phase combinations in Equations (24) should be considered as the fast phase variables. Averaging over these fast phase variables yields the following system of the slow-flow equations:

$$\dot{J} = -\frac{F}{2} \left( q_1(J) e^{i\vartheta} + q_1^*(J) e^{-i\vartheta} \right)$$

$$\dot{\vartheta} = \frac{\partial h_0(J)}{\partial J} - \frac{iF}{2} \left( \frac{\partial q_1(J)}{\partial J} e^{i\vartheta} - \frac{\partial q_1^*(J)}{\partial J} e^{-i\vartheta} \right) - \Omega$$

(25)

Here $J(t) = \langle I(t) \rangle$ is the average of the action variable over the fast phases. It is easy to reveal by direct differentiation that System (25) possesses the following first integral:

$$C = h_0(J) - \frac{iF}{2} \left( q_1(J) e^{i\vartheta} - q_1^*(J) e^{-i\vartheta} \right) - \Omega J = \text{const}.$$

(26)

Expression (26) defines a family of 1:1 RMs of the system. The constant $C$ is determined by the IC on the RM– the values of the action and the slow phase, at which the system is captured by the RM. The first integral (27) is a particular case of general conservation law for the RMs of the single-DOF systems with periodically time-dependent Hamiltonian [29].

For particular case of the perturbed truncated well $U_\varepsilon(q)$ it is convenient to choose the basic Hamiltonian of the unperturbed well (linear oscillator) and to use regular AA transformation:

$$H_b = \frac{p^2 + q^2}{2}, q = \sqrt{2I} \sin \theta, p = \sqrt{2I} \cos \theta$$

$$H_0 = H_b - \frac{\varepsilon \alpha q^4}{4} = I - \varepsilon \alpha I^2 \sin^4 \theta \tag{27}$$

Therefore, the integral of motion (26) for this system takes the following form:

$$C = J - \frac{3\varepsilon \alpha}{8} J^2 + F \sqrt{\frac{J}{2}} \cos \vartheta - \Omega J = \text{const} \tag{28}$$

In order to explore the perturbation of the main resonance dip, we formally consider the case of small forcing $F = \varepsilon f, f \sim O(1)$ and small frequency detuning $\Omega = 1 + \varepsilon \sigma, \sigma \sim O(1)$. Then, the conservation law (28) is simplified to the following:

$$D = -\frac{3\alpha}{8} J^2 + f \sqrt{\frac{J}{2}} \cos \vartheta - \sigma J \tag{29}$$

Conservation law (29) does not include the bookkeeping parameter $\varepsilon$ and the forcing phase $\Psi$. The latter observation corroborates with Equation (17) the dip shape for the unperturbed problem also does not depend on $\Psi$ in the main approximation. Physically, this result means that the averaging over the fast time scale suppresses the dependence on the forcing phase.

For convenience we denote $J = 2N^2$ and obtain

$$D = -\frac{3\alpha}{2} N^4 + fN \cos \vartheta - 2\sigma N^2 \tag{30}$$

This expression is used for further exploration of the transient dynamics on the RM. The phase portrait on the RM corresponds to the level set of conservation law (30). In accordance with (4) and (27), the escape event in the averaged system should be associated with phase trajectory approaching the limit

$$N_{\max} = 1/2 \tag{31}$$

for some value of the slow phase $\vartheta$. Thus, the particle will escape the well if the RM trajectory that corresponds to selected IC will achieve the value of $N_{\max}$. As it was already mentioned above, in current work we selected zero and thus will refer to this special trajectory as limiting phase trajectory (LPT), in accordance with some recent studies on transient dynamics in nonlinear oscillatory systems [23, 24, 30, 31]. From expression (30) it follows that the LPT is determined by condition $D=0$. We note in passing that there is nothing special about the LPT in current problem, and for different IC, the escape would be governed by other phase trajectory on the RM.

One can identify two main phase mechanisms, or scenarios, of transition to the escape on the RM. The first mechanism reveals itself for positive and small negative values of the detuning, and is illustrated in Figure 4.

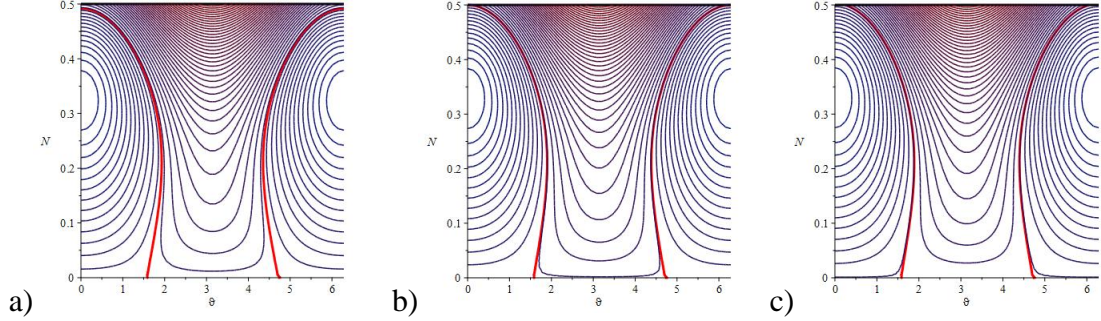

*Figure 4. Structure of the RM for positive and small negative values of the detuning. Red line denotes the LPT ($D=0$), thick black line corresponds to $N = N_{max} = 1/2$. $\sigma = -0.1, \alpha = 1$ a) $f = 0.08$; b) $f = f_{cr} = 0.0875$; c) $f = 0.09$.*

In this simple scenario, for $f < f_{cr}$ the LPT does not achieve $N_{max}$, and the particle remains in the well (Figure 4a). For $f > f_{cr}$ the LPT achieves $N_{max}$, and the particle escapes (Figure 4c). At the boundary $f = f_{cr}$ the line $N = N_{max}$ is tangent to the LPT at $\vartheta = 0$. We will refer to this transition as Scenario I.

For larger negative values of the detuning, the transition to escape has more complicated scenario, as demonstrated in Figure 5.

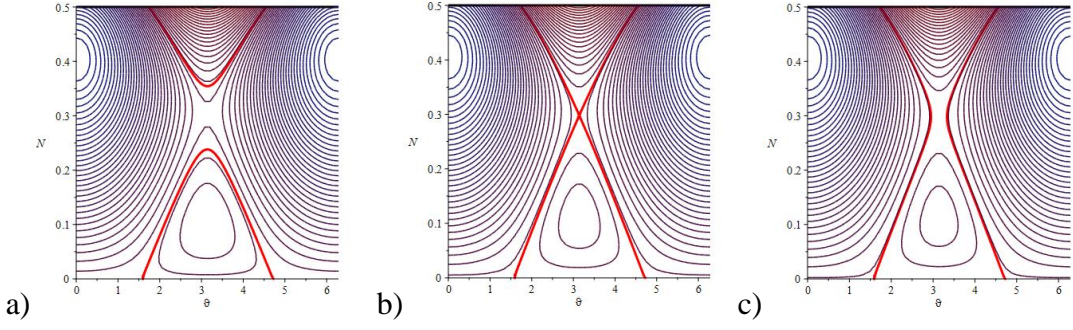

a) b) c)

*Figure 5. Structure of the RM for large negative values of the detuning. Red line denotes the LPT ($D = 0$), thick black line corresponds to $N = N_{max} = 1/2$. $\sigma = -0.2, \alpha = 1$ a) $f = 0.075$; b) $f_{cr} = 0.079504$; c) $f = 0.081$.*

In this scenario, for $f < f_{cr}$ the LPT remains below the saddle point on the RM, and the particle stays in the well. For $f > f_{cr}$ the escape occurs. The boundary $f = f_{cr}$ corresponds to passage of the LPT through the saddle point at the RM. This transition will be referred to as Scenario II.

Due to relative simplicity of the integral of motion (30) it is easy to obtain explicit expressions for dependences of critical forcing on the detuning for both transition scenarios described above. For Scenario I at $f = f_{cr}$ the LPT achieves $N_{max}$ at $\vartheta = 0$, therefore, one obtains:

$$D(f = f_{cr}, N = 1/2, \vartheta = 0) = 0 \Rightarrow f_{cr\_I} = \frac{3\alpha}{16} + \sigma \qquad (32)$$

For the saddle in point $(\vartheta, N) = (\pi, N_s)$ the following relationship should be satisfied:

$$\left.\frac{\partial D(\vartheta, N)}{\partial N}\right|_{\vartheta=\pi, N=N_s} = 0 \Rightarrow f + 4\sigma N_s + 6\alpha N_s^3 = 0 \qquad (33)$$

As for Scenario II, at $f = f_{cr}$ the LPT passes through the saddle point:

$$D(f = f_{cr}, \vartheta = \pi, N = N_s) = -\frac{3\alpha}{2} N_s^4 - f_{cr} N_s - 2\sigma N_s^2 = 0 \qquad (34)$$

Excluding $N_s$ from (33) and (34), one obtains for Scenario II:

$$f_{cr\_II} = \frac{8}{9} \frac{(-\sigma)^{3/2}}{\alpha^{1/2}} \tag{35}$$

In Figure 6 we depict the theoretical prediction for the escape threshold, based on expressions (32,35).

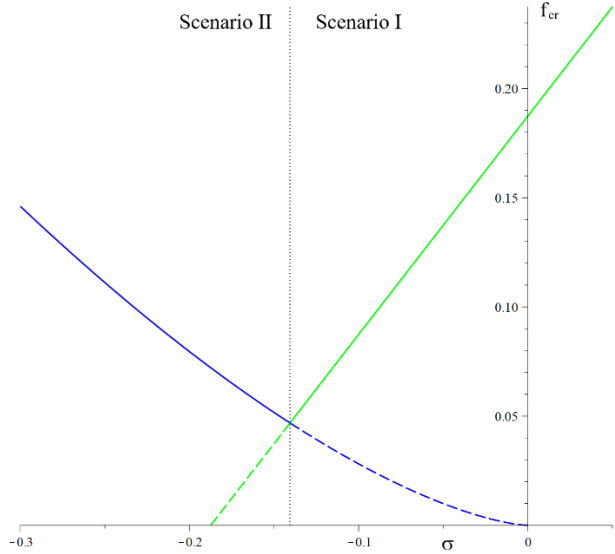

*Figure 6. Theoretical prediction for the escape threshold. Green line corresponds to expression (32), blue line – to expression (35). Vertical dotted line divides between different escape scenarios on the RM. Dashed parts of the curves mean that the corresponding escape scenario is "overruled" by the alternative one.*

Solid lines in Figure 6 form the theoretical escape threshold, in a form of sharp dip. Coordinates of the minimum $(\sigma_*, f_*)$ easily follow from (32,35) and are written as follows:

$$\sigma_* = \frac{-9\alpha}{64}, \quad f_* = \frac{3\alpha}{64} \tag{36}$$

One readily observes that the sharp minimum (36) has nonzero forcing amplitude and occurs for negative value of the detuning (i.e. below the natural frequency of the oscillator). Moreover, as the perturbation disappears ($\alpha \to 0$) the minimum tends to (0,0) point. Thus, it is possible to conclude that the generic symmetric perturbation of the truncated parabolic well shifts the sharp dip towards lower frequencies and higher amplitudes, in complete agreement with the phenomenological observations for various model systems.

In Figure 7 we present the RM phase portrait that corresponds to the minimum point $(\sigma_*, f_*)$.

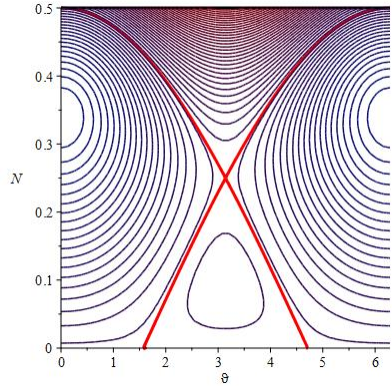

*Figure 7. Phase portrait of the RM for minimum point $(\sigma_*, f_*)$.*

In this case, the LPT is simultaneously homoclinic to the saddle point $(\pi, N_s)$ and tangent to $N = N_{\max}$. In other terms, two escape scenarios merge together, as expected.

*3.3 Numeric verifications.*

Approximate character of the treatment presented above calls for numeric verification of the results. First, we demonstrate that the transition to the escape indeed can occur by two qualitatively different scenarios, by direct numeric simulation of Equation (19). The simulation results for the Scenario I are presented in Figure 8, and for the Scenario II – in Figure 9.

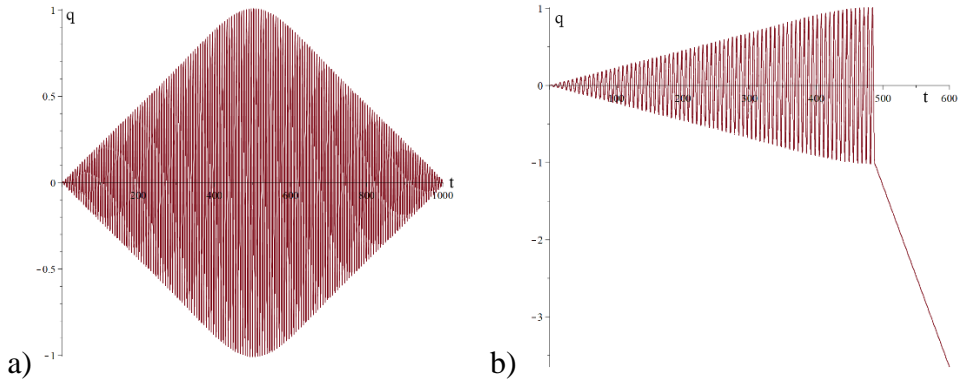

Figure 8. Time series for Equation (19) with zero initial conditions, $\varepsilon = 0.05$, $\alpha = 1$, $\sigma = -0.1$, a) $f = 0.09$; b) $f = 0.093$

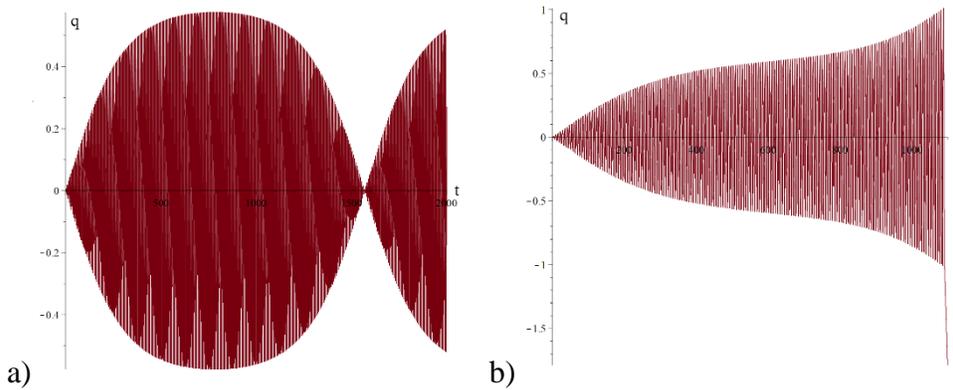

Figure 9. Time series for Equation (19) with zero initial conditions, $\varepsilon = 0.05$, $\alpha = 1$, $\sigma = -0.2$, a) $f = 0.079$; b) $f = 0.080$

In Figure 8a the modulation envelope almost achieves the boundary of the well, and for slightly higher forcing amplitude (Figure 8b) the particle escapes. Such behavior is typical for Scenario I. In the same time, In Figure 9a the modulation envelope achieves the maximum of about 0.49 – quite far from the well boundary. However, also in this case for just slightly higher forcing amplitude the particle indeed escapes (Figure 9b). Also, one notices substantial difference between the modulation envelopes in Figures 8a and 9a. Slow evolution of the envelope near the maximum is caused by proximity of the LPT to the saddle point at the RM. One can also trace a similar effect of slowing the modulation dynamics close to amplitude 0.5 – also due to proximity of the saddle. So, in this case the escape transition is governed by the RM saddle point, in complete agreement with Scenario II.

In order to assess the accuracy of predictions (32) and (35) for the escape thresholds, we note that it is possible to re-write these predictions in rescaled form:

$$\frac{f_{cr\_I}}{\alpha} = \frac{3}{16} + \frac{\sigma}{\alpha}; \quad \frac{f_{cr\_II}}{\alpha} = \frac{8}{9}\left(-\frac{\sigma}{\alpha}\right)^{3/2} \tag{37}$$

In accordance with (37), properly rescaled data for different values of $\alpha$ should collapse on the same plot. This prediction is verified in Figure 10.

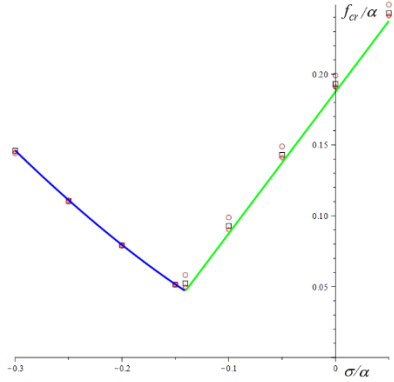

Figure 10. Collapse of numeric escape threshold in coordinates $(\sigma/\alpha, f_{cr}/\alpha)$, $\varepsilon = 0.05$, $\Psi = 0$, diamonds - $\alpha = 0.5$, boxes - $\alpha = 1$, circles - $\alpha = 2$. Blue line - $\frac{f_{cr}}{\alpha} = \frac{8}{9}(-\sigma/\alpha)^{3/2}$, $-0.3 < \sigma/\alpha < -9/64$; green line - $\frac{f_{cr}}{\alpha} = 3/16 + \sigma/\alpha$, $-9/64 < \sigma/\alpha < 0.05$

The data collapse almost perfectly, especially for the left branch that describes the escape scenario with the saddle point. Therefore, in current model the theoretical evaluation nicely predicts the observed escape threshold. This success may be attributed to quasilinear interactions *inside* the considered potential well – thus, application of the perturbative technique is well justified. In the same time, the complete model should be considered as essentially nonlinear, due to the truncation of the potential well. In earlier works devoted to the escape, smooth quartic potential has been considered [4]. In the next Section we are going explore the transient escape dynamics for this benchmark model.

## 4. Escape dynamics in $\varphi^4$ model.

Here we pass to "genuine" essentially nonlinear potential well and consider the complete $\varphi^4$ potential with negative nonlinearity. Hamiltonian of this system is written in the following form:

$$H = \frac{p^2}{2} + \frac{q^2}{2} - \frac{q^4}{4} - Fq\sin(\Omega\tau + \Psi) \tag{38}$$

All coefficients that are set to unity can be scaled out without reducing the generality. Equation of motion that corresponds to Hamiltonian (38) has the form

$$\ddot{q} + q - q^3 = F\sin(\Omega\tau + \Psi) \tag{39}$$

In this case, the nonlinearity is substantial, and it seems not possible to rely on quasilinear approach developed in the previous Section. Still, one can apply the averaging procedure outlined above in (22-26). However, the basic Hamiltonian for the AA transformation is selected as *complete* Hamiltonian of the unforced φ⁴-oscillator:

$$H_b(p,q) = H_0(p,q) = \frac{p^2}{2} + \frac{q^2}{2} - \frac{q^4}{4}, \quad -1 \leq q \leq 1 \tag{40}$$

Transformation to the AA variables for this Hamiltonian is available in literature, but for the sake of completeness, we present a brief derivation in the Appendix. The resulting transformation is the following:

$$I(E) = \frac{2\sqrt{2}}{3\pi}\sqrt{1+\mu}(\mathbf{E}(k) - \mu\mathbf{K}(k)); \quad \mu = \sqrt{1-4E}, \quad k = \sqrt{\frac{1-\mu}{1+\mu}}$$
$$q(\theta) = \sqrt{1-\mu}\operatorname{sn}\left(\frac{2\mathbf{K}(k)}{\pi}\theta, k\right) \tag{41}$$

Here $E = H_0(p,q) = \text{const}$ is the energy of the unforced $\varphi^4$ oscillator, $\mathbf{K}(k)$ and $\mathbf{E}(k)$ are complete elliptic integrals of the first and the second kind respectively, $k$ is the modulus of elliptic functions, and $\text{sn}(x,k)$ is Jacobi elliptic sine. Expression (41) is rather complicated and definitely cannot be inverted to obtain explicit dependence $E(I)$. This obstacle, however, does not disqualify the derivation (22-26), but makes it necessary to parametrize the integral of averaged motion of the forced system (26) via $(\xi, \vartheta)$ instead of $(J, \vartheta)$. Here $\xi(t) = \langle E(t) \rangle$ is the energy of forced oscillator averaged over the fast time scale. Besides, we use well-known nomal expansion of the Jacobi elliptic sine into Fourier series:

$$\text{sn}\left(\frac{2\mathbf{K}(k)}{\pi}\theta, k\right) = \frac{2\pi}{\mathbf{K}(k)k} \sum_{n=0}^{\infty} \frac{Q^{n+1/2}}{1-Q^{2n+1}} \sin((2n+1)\theta)$$
$$Q = \exp\left(-\frac{\pi \mathbf{K}'(k)}{\mathbf{K}(k)}\right), \quad \mathbf{K}'(k) = \mathbf{K}(\sqrt{1-k^2}) \tag{42}$$

In fact, we need only the first term from expansion (42). Explicitly the integral of averaged motion of system (38) is written in the following form:

$$C = \xi - FG(\xi)\cos\vartheta - \Omega J(\xi) = \text{const}$$
$$\mu(\xi) = \sqrt{1-4\xi}, \; k(\xi) = \sqrt{\frac{1-\mu(\xi)}{1+\mu(\xi)}}, \; J(\xi) = \frac{2\sqrt{2}}{3\pi}\sqrt{1+\mu(\xi)}(\mathbf{E}(k(\xi)) - \mu(\xi)\mathbf{K}(k(\xi))) \tag{43}$$
$$G(\xi) = \frac{\pi\sqrt{1-\mu(\xi)}}{\mathbf{K}(k(\xi))k(\xi)} \exp\left(-\frac{\pi \mathbf{K}'(k(\xi))}{2\mathbf{K}(k(\xi))}\right)$$

Expression (43) determines the phase flow on the 1:1 RM for System (38). The natural escape criterion for this system is approaching the boundary value of the energy $\xi = 0.25$. Zero IC force the choice of phase trajectory (LPT) corresponding to $C = 0$. Similarly to Section 3, one can identify two different mechanisms for transition to the escape for large and small excitation frequencies respectively. The mechanisms are illustrated in Figures 11 and 12.

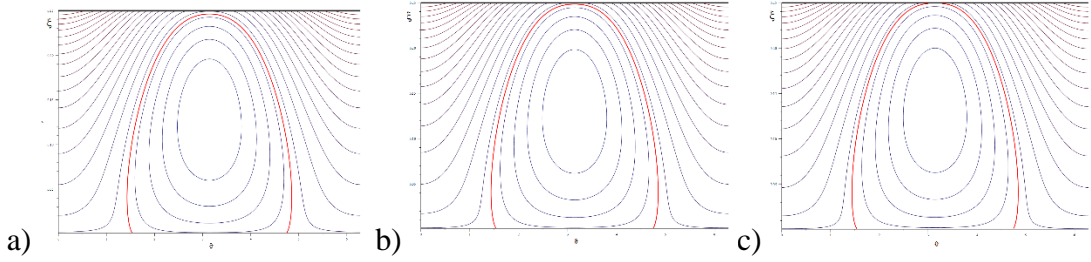

*Figure 11. Structure of the RM defined by Equation (43) for the case of large forcing frequencies. Red line denotes the LPT ( $C = 0$ ), thick black line corresponds to $\xi = 1/4$. $\Omega = 0.95$  a) $F = 0.068$; b) $F_{cr} = 0.071$; c) $F = 0.073$.*

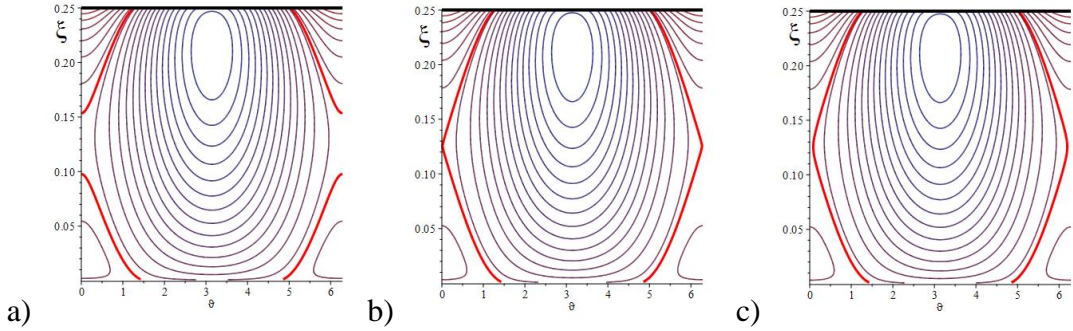

*Figure 12. Structure of the RM defined by Equation (43) for the case of small forcing frequencies. Red line denotes the LPT ( $C = 0$ ), thick black line corresponds to $\xi = 1/4$. $\Omega = 0.8$  a) $F = 0.069$; b) $F_{cr} = 0.07075$; c) $F = 0.071$.*

Two different mechanisms of the escape transition, similar to Scenarios I and II described in the previous Section, clearly reveal themselves also in this case. Moreover, the two scenarios merge together at the point of minimal possible forcing for the escape at given IC, as presented in Figure 13.

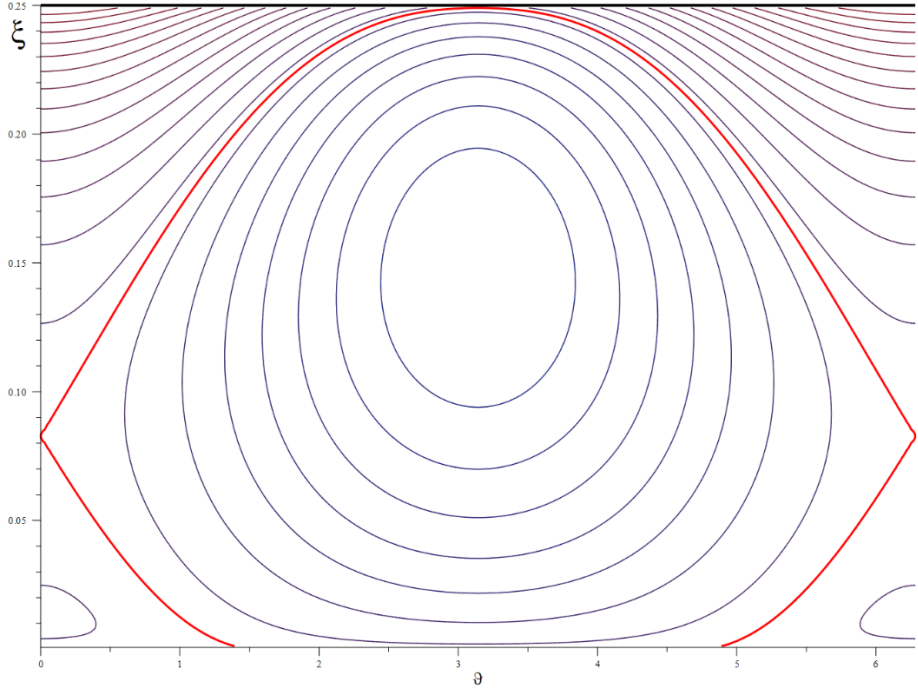

*Figure 13. Structure of the RM defined by Equation (43) for the minimum at frequency-critical forcing plot. Red line denotes the LPT ($C = 0$), thick black line corresponds to $\xi = 1/4$. $\Omega_* = 0.886427, F_* = 0.0318186$*

To derive the equations for the escape thresholds according to Scenarios I and II, we use integral of motion (43). The escape threshold for Scenario I is described by straight line on $\Omega - F_{cr}$ plane:

$$C(\vartheta = \pi, \xi = \xi_{max}) = 0 \Rightarrow \xi_{max} + F_{cr}G(\xi_{max}) - \Omega J(\xi_{max}) = 0 \tag{44}$$

Here $F_{cr}$ denotes the minimal forcing amplitude required for the escape for given forcing frequency and IC. For computing coefficients in Equation (43) we use $\xi_{max} = 0.249$ due to logarithmic singularity in $G(\xi)$.

For Scenario II, the LPT should pass through the saddle point at $\vartheta = \pi$. Then, one obtains two equations:

$$\begin{aligned} F_{cr}G(\xi) + \Omega J(\xi) &= \xi \\ F_{cr}G'(\xi) + \Omega J'(\xi) &= 1 \end{aligned} \tag{45}$$

Equations (45) determine the line $F_{cr}(\Omega)$ in parametric form. Corresponding plots with superimposed results of direct numeric simulation are presented in Figure 14.

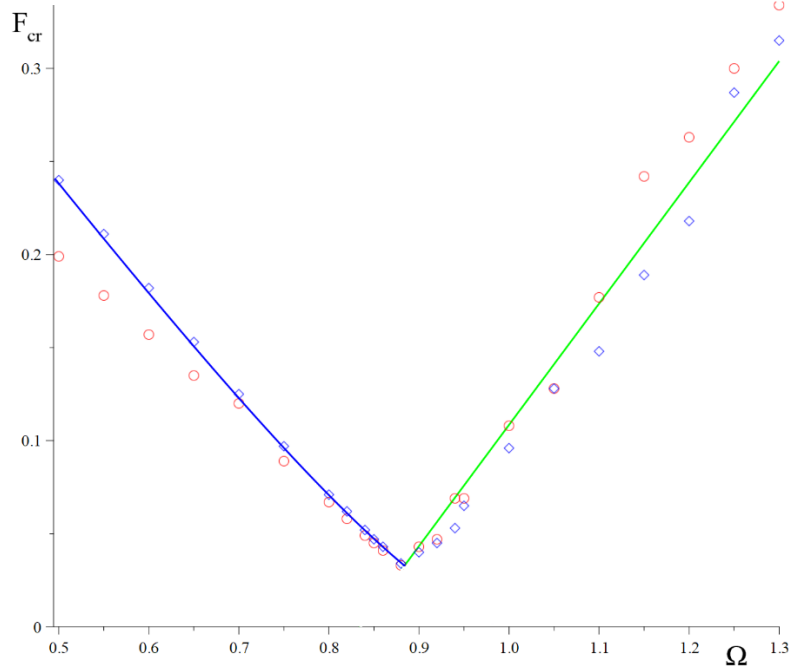

*Figure 14. Comparison of theoretical and numeric results for the escape thresholds. Green line corresponds to Scenario I (Figure 11, Equation (44)), blue line – to Scenario II (Figure 12, Equations (45)). Blue diamonds correspond to $\Psi = 0$, red circles – to $\Psi = \pi/2$*

Despite lack of formal small parameter in Hamiltonian (38), one observes that the agreement between the results of primary averaging and the results of direct numeric simulation is quite satisfactory. The best accuracy is achieved in the most interesting region near the minimum. Presumably, this happens since the value of the critical forcing in this case is rather small, and the forcing itself serves as hidden small parameter for the averaging procedure.

5. **Conclusions**

The results presented in the paper relate the observed universal phenomenology of the forcing-frequency curves in various escape models to simple and easily tractable case of the harmonically forced and undamped particle inside the truncated parabolic well. More exactly, one can say that the observed sharp minimum (dip) of the critical forcing is "generated" by the point of exact 1:1 resonance, in which the critical force tends to zero. Added soft nonlinearity results in the shift of the dip towards nonzero minimal critical forcing at certain frequency below the natural frequency of the well. Averaging procedure allows complete analysis of the transient dynamics in the perturbed case, and delivers very accurate prediction of the escape threshold. Moreover, it was demonstrated that "complete" strongly nonlinear $\varphi^4$ model demonstrates very similar escape patterns on the resonance manifold. Also for this model, the escape thresholds near the minimum can be predicted with the help of averaging procedure based on the action-angle transformation. Very good accuracy is achieved without any correction factors and/or *ad hoc* escape criteria used in some earlier works.

The results also demonstrated that, similarly to earlier treatment [22], the sharp minimum is formed due to competition between two escape mechanisms on the RM. In the same time, the hypothesis in [22] concerning universality of these mechanisms has been disproved. New mechanism revealed above involves achieving the escape boundary by the LPT without interfering with saddle points. Thus, one has to conclude that specific escape mechanisms on the RM are model-dependent.

Many issues mentioned in passing above remain unexplored. For instance, the results on the truncated parabolic well clearly indicate that beyond the main resonance the forcing phase has strong effect on the escape threshold. Besides, it was demonstrated that in certain resonance points the threshold forcing appears to be substantially *higher* than the universal estimation (8). In the same time, it is well-known that the forcing –frequency dependence demonstrates *additional dips* in the vicinity of these resonances [4]. This apparent contradiction calls for further clarification. Other open issue is the effect of damping. For some particular model, it is already known [25] that the damping also shifts the minimum towards higher amplitudes and lower frequencies, while preserving qualitatively similar escape mechanisms at the RM. Still, in general this problem is left for future research.

**Funding**

The authors are very grateful to Israel Science Foundation (grant 1696/17) for financial support of this work.

**Conflict of interests**

The authors declare that they have no conflict of interests

## 6. Appendix

In this Appendix, we present the derivation of the action-angle transformations (41) for the $\varphi^4$ potential with softening nonlinearity.

We start from the unperturbed Hamiltonian (40):

$$H_b(p,q) = \frac{p^2}{2} + \frac{q^2}{2} - \frac{q^4}{4} = E = \text{const}, \quad -1 \leq q \leq 1 \tag{A1}$$

According to Equation (22), and taking into account the symmetry considerations, the expression for action is written as follows:

$$2\pi I(E) = 4\int_0^{q_{max}} \sqrt{2E - q^2 + \frac{q^4}{2}}\,dq = 2\sqrt{2}\int_0^{q_{max}} \sqrt{(1-q^2)^2 - (1-4E)}\,dq \quad (A2)$$

$$q_{max} = \sqrt{1-\mu},\ \mu = \sqrt{1-4E}$$

Further change of variables $1 - q^2 = z$ yields:

$$2\sqrt{2}\int_0^{q_{max}} \sqrt{(1-q^2)^2 - (1-4E)}\,dq =$$

$$= \sqrt{2}\int_\mu^1 \sqrt{\frac{z^2 - \mu^2}{1-z}}\,dz = \frac{4\sqrt{2}}{3}\sqrt{1+\mu}(\mathbf{E}(k) - \mu\mathbf{K}(k)),\ k = \sqrt{\frac{1-\mu}{1+\mu}} \quad (A2)$$

Angle variable is expressed as follows:

$$\theta = \frac{\partial}{\partial I}\int_0^q \sqrt{2E - q^2 + \frac{q^4}{2}}\,dq = \frac{dE}{dI}\int_0^q \sqrt{\frac{1}{(1-q^2)^2 - (1-4E)}}\,dq = \frac{\pi}{2\mathbf{K}(k)}\mathbf{F}(\varphi, k)$$

$$\varphi = \arcsin\sqrt{\frac{1-z}{1-\mu}} = \arcsin\frac{q(\theta)}{\sqrt{1-\mu}} \quad (A3)$$

Here $\mathbf{F}(\varphi, k)$ is the incomplete elliptic integral of the first kind. Expressions (41) easily follow from (A1-A3).